# High-Throughput Screening of Transition Metal Single-Atom Catalysts for Nitrogen Reduction Reaction

Tingting Bo[1], Shiqian Cao[1], Nan Mu[1], Ruixin Xu[1], Yanyu Liu[2]*, Wei Zhou[1]*

1. Department of Applied Physics, Tianjin Key Laboratory of Low Dimensional Materials Physics and Preparing Technology, School of Science, Tianjin University, Tianjin 300072, P. R. China.
2. College of Physics and Materials Science, Tianjin Normal University, Tianjin, 300387, P. R. China.
Corresponding Author: *E-mail: yyliu@tjnu.edu.cn; weizhou@tju.edu.cn

**Abstract:** The discovery of metals as catalytic centers for nitrogen reduction reactions has stimulated great enthusiasm for single-atom catalysts. However, the poor activity and low selectivity of available SACs are far away from the industrial requirement. Through the high throughout first principles calculations, the doping engineering can effectively regulate the NRR performance of $\beta$-Sb monolayer. Especially, the origin of activated $N_2$ is revealed from the perspective of the electronic structure of the active center. Among the 24 transition metal dopants, Re@Sb and Tc@Sb showed the best NRR catalytic performance with a low limiting potential. The Re@Sb and Tc@Sb also could significantly inhibit HER and achieve a high theoretical Faradaic efficiency of 100%. Our findings not only accelerate discovery of catalysts for ammonia synthesis but also contribute to further elucidate the structure‐performance correlations.

# 1. Introduction

The synthesis of NH$_3$ is crucial for the industrial production, including fertilizer, chemical production and so on. At present, it is produced mainly by the traditional Haber−Bosch reaction, that is, the transformation of N$_2$ and H$_2$ to NH$_3$ was successfully achieved under high temperature and high pressure.[1] However, after hundreds of years of development, its conversion efficiency is low, accompanied by huge energy consumption and the release of CO$_2$, aggravating energy and environmental problems.[2-4] Therefore, it is urgent to find green ammonia synthesis technology. The electrochemical nitrogen reduction reaction (eNRR) using water as hydrogen source under ambient conditions has become the most promising alternative to Haber−Bosch reaction.[5-8] The electrocatalytic NRR has mild reaction conditions and simple operation. In addition, water is used as raw material to provide the hydrogen source required by NRR through hydrogen evolution reaction (HER), breaking away from the heavy dependence on fossil fuels of the Hubble process. Hydrogen is produced as a sustainable proton source by electrochemical splitting of water through hydrogen evolution. This method can reduce the restriction of thermodynamic equilibrium and control the direction and rate of reaction by changing the external field. In recent years, metals, metal oxides and transition metal chalcogenides have been extensively applied in NRR. However, the N$_2$ activation and low Faraday efficiency are two major challenges for NRR.[9, 10] Therefore, it is more important to search for eNRR catalysts with high-efficiency, low-cost and environmental friendliness.

Usually, chemisorption on the catalyst surface is difficult due to the inertness of the N≡N triple bond, which will result in high overpotential and low yield of the NRR process.[11] In order to improve NRR activity, the catalysts need to enhance chemisorption of N$_2$ to ensure fully activation of the N≡N triple bond. Generally speaking, the strong adsorption of N$_2$ is favorable to the activation of N$_2$, but this also indicates that the reaction product NH$_3$ is strongly adsorbed on the catalyst surface, which will lead to the failure of rapid desorption of NH$_3$.[12] Therefore, an excellent catalyst needs to balance the adsorption of $N_2$ and desorption of $NH_3$. Moreover, the HER is the most important competitive reaction for NRR. That is to say, the catalyst selectively adsorbed H before $N_2$ molecules, resulting in the blocking of the active site and reducing the Faraday efficiency of $NH_3$ production.[13]

The single atom catalysts (SACs), as a new type of heterogeneous catalysts, have attracted extensive attention. Specifically, compared with metal surface catalysis, single atom catalysis has clear active center, high selectivity, maximum atomic efficiency and good stability.[14-16] This has been proved by numerous theoretical and experimental studies. Theoretically, Li *et al*. proposed the reduction of $NH_3$ by $N_2$ in Mo embedded at the defect of graphene for the first time at room temperature and pressure.[17] Many SACs are predicted to be high-performance NRR catalysts, including Nb-graphene,[18] Ru-G-$C_3N_4$,[19] and single-boron catalysts[20] and so on. In most SACs, anchoring or doping of transition metals on two-dimensional material surfaces predominate in promising NRR catalysts, because the activation of $N_2$ is related to the arrangement of the TM $d$ orbital.[21] The $d$ orbitals of TM have empty orbitals that can accept the electrons in the $\sigma$ bond of $N_2$, and occupied electrons that can fill the antibonding orbitals of $N_2$, which will strengthen the TM-N bond and weaken the N≡N triple bond, respectively.[22, 23]

On the other hand, two-dimensional (2D) materials have been active in catalysis research fields, including the nitrogen reduction reaction,[24] hydrogen evolution reaction,[25] $CO_2$ reduction reaction ($CO_2$RR),[26] and oxygen reduction reaction (ORR),[27] due to the high surface area and more abundant exposed active sites. In recent years, group-VA monolayer semiconductor materials ($\beta$-Sb), a new two-dimensional material, has been theoretically predicted and then experimentally synthesized.[28, 29] In particular, with the unique structure and exceptional electronic properties, it is widely used in various fields, including optoelectronic,[30] sensing devices,[31] energy devices and so on.[32, 33]

Along with experimental attempts, theoretical screening and designing based on DFT simulations are also essential to promote NRR catalysis. In this work, we investigate the NRR activity of a series of TM doped β-Sb by first-principles calculations. Specifically, we screened 24 TM@Sb NRR catalysts candidates one by one through three screening criteria: stability, catalytic activity and selectivity. The screening results show that the NRR limiting potential of Ti, Tc, Mo, Nb and Re doped β-Sb is lower than that of ideal NRR catalyst Ru (-0.55 V). In particular, the Re@Sb is potential optimal NRR catalysts with extremely low NRR limiting potential and good HER competitiveness. This work not only provides a complete framework and screening criteria for NRR of Sb catalysts, but also provides preliminary theoretical guidance for future experiments.

## 2. Computational model and details

In this paper, all calculations were performed by the Vienna Ab initio Simulation Package (VASP) based on density functional theory (DFT).[34, 35] Via generalized gradient approximation (GGA), the exchange correlation functional was implemented by the Perdew–Burke–Ernzerhof (PBE) functional.[36] The Grimme method (DFT-D2) was used to describe the weak van der Waals interactions.[37] We also adopted the projector augmented wave potentials (PAW) to describe ion-electron interactions. The cutoff energy is set as 400 eV. In addition, we tested the value of cutoff energy. Taking Re@Sb adsorption $N_2$ as an example, as shown in Figure S1, when the cutoff energy is 400 eV, the energy change is about 0.001 eV/atom, and the energy parameter converges. The crystal structures were completely optimized until the convergence criterion of total energy and residual force are $10^{-5}$ eV and 0.01 eV/Å, respectively. Furthermore, a gamma-centered scheme of 2 ×2 ×1 $k$-point mesh for the 3 ×3 ×1 supercell was used to sample the corresponding Brillouin zone. In addition, we use an 18 Å vacuum perpendicular to the surface to avoid adjacent interactions caused by periodic models. The spin polarization is turned on to take account of the magnetic presence of the system. Bader charge analysis was used to analyze the atomic charges and charge transfer. The details of Gibbs free energy, binding energy and Faraday efficiency calculations are given in the supporting information.

In addition, in order to better evaluate the catalytic performance, the quantum transport properties of catalysts were performed using NANOD-CAL package.[38, 39] Nanodcal is a code based on non-equilibrium Green's function density functional theory (NEGF-DFT). It is mainly used to simulate the nonlinear and non-equilibrium quantum transport processes in materials. The device model constructed is shown in Figure S2. The simulated device is divided into three regions: the left electrode, the right electrode and central scattering region. In detail, the length of left/right electrode and scattering region are summarized in Table S4. Moreover, the transport direction is along the $x$ axis and the vacuum layer is set to 18 Å along the $z$ axis to avoid interlayer effect. In the self-consistent calculations, a $1 \times 4 \times 1$ $k$-point mesh is used, with the convergence criterion set to less than $10^{-4}$ Hartree for the density matrix of every element. For the central region $k$-space integration, a $1 \times 100 \times 1$ $k$-point mesh is used to calculate the current-voltage (I-V) curve and transmission spectrum.

## 3. Results and discussion

### 3.1 The geometries and stabilities of SACs.

The $\beta$-Sb monolayer, a layered rhomboid bilayer structure composed of interlocking, folding and six-membered rings, has attracted more and more attention in the field of catalysis. Furthermore, the optimized lattice parameter of $\beta$-Sb monolayer is $a = b = 4.14$ Å, which is consistent with previous experiments and theories investigations.[29, 40] In order to study the NRR performance of Sb-based catalysts systematically, we selected a single transition metal atom for catalytic doping in $\beta$-Sb monolayer, TM@Sb, where TM is one of the 3$d$, 4$d$ or 5$d$ transition metal atoms listed in Figure 1b. Regrettably, the monolayer geometrical structure is greatly deformed after complete structural relaxation when doped with IB and IIB elements, so it will not be considered in subsequent studies.

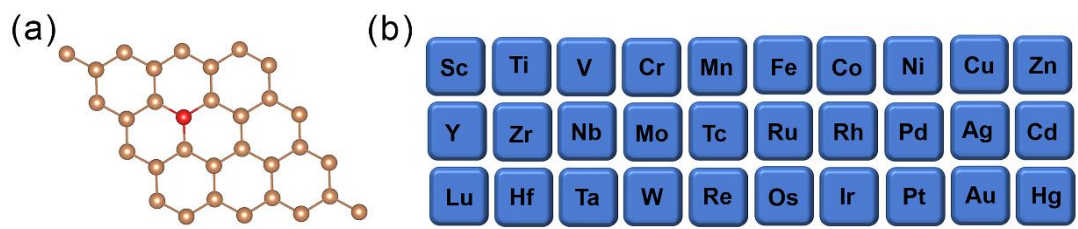

Figure 1. (a) The geometrical structure of TM atom doped SAC on $\beta$-Sb monolayer. (b) list of metals utilized in this study.

In this section, screening criteria were used to describe the catalytic performance of these active sites. As Figure 2a shows, first of all, the single TM atom doped $\beta$-Sb system satisfies the thermodynamic stability ($E_b < 0$). Then, N$_2$ should be chemisorbed and activated efficiently with its corresponding change of Gibbs free energy with $\Delta G_{*N2}$ < -0.30 eV. Third, in order to reduce the energy cost of the N$_2$ reduction reaction, the Gibbs free energy change of the first ($\Delta G_{*N2-N2H}$) to be less than 0.55 eV, which is the limiting potential of the most considered best NRR catalyst Ru. Afterward, regarding potential candidates, the Gibbs free energy change of the potential-determining step (PDS) to be less than 0.55 eV. Lastly, for better selectivity of NRR, the maximum free energy change of NRR should be smaller than the competing HER ($\Delta G_{*N2} < \Delta G_{*H}$).

The good stability is an important guarantee to maintain effective catalysis of SACs. Therefore, we calculated the binding energy of TM doped system to examine the stability of SACs, as shown in Figure 2b. In general, the negative $E_b$ implies that TM atoms prefer stable doping in Sb monolayers. Clearly, all of the SACs can satisfy the criteria. The more negative binding energy means that the vacancy in Sb is an ideal anchor site, ensuring high stability.

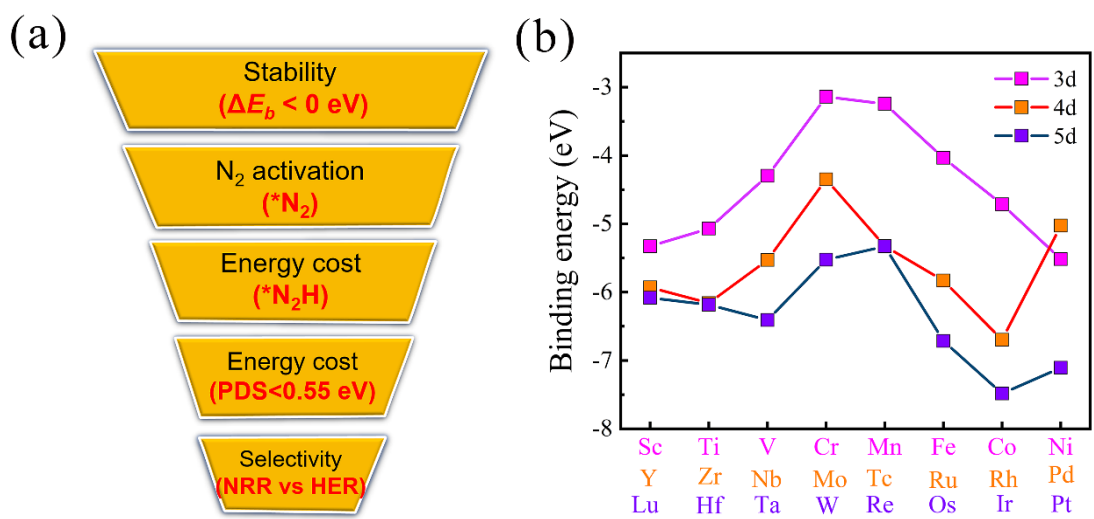

Figure 2. (a) Proposed "Five-Step" strategy for screening NRR candidate catalysts. (b) Computed $\Delta E_b$ of metal atoms doped on $\beta$-Sb monolayer.

## 3.2 The adsorption and activation of the $N_2$ molecule

The adsorption and activation of $N_2$ is the most important process of NRR reaction, which will affect the overall reaction efficiency, because $N_2$ adsorption is the basic premise for starting the subsequent hydrogenation process.[41] In general, the three possible adsorption configurations of $N_2$ on the catalyst consists of chemisorption via end on or side on configuration and physisorption, as shown in Figure 3a. The Gibbs free energy of the adsorbed $*N_2$ ($\Delta G_{*N2}$) were considered to effective descriptors of $N_2$ capture and activation. Here, the value of $\Delta G_{*N2}$ less than -0.30 eV indicates that the adsorption of $N_2$ by SACs is a thermodynamic spontaneous reaction and activated $N_2$ effectively. On the contrary, the $\Delta G_{*N2}$ values greater than -0.30 eV indicate that $N_2$ adsorption and activation hardly occur at room temperature.

Therefore, the $\Delta G_{*N2}$ greater than -0.30 eV is not further discussed. First, $N_2$ chemical adsorption in both end on and side on configurations on the pristine $\beta$-Sb monolayer is considered. Since every Sb atom in pristine structure has a perfect $sp^3$ hybridization and octet electron configuration, they have no chemical activity for $N_2$ adsorption. Therefore, the influence of doping TM on the activity of the adsorption site was discussed, and its optimized representative geometric structure is shown in Figure

S3 (relative to free $N_2$ and isolated Sb monolayer). The Gibbs free energies of *$N_2$ is presented in Figure 3b, respectively. In addition, part of the side-on adsorption model was converted to the standing adsorption, as shown in Figure S4, so the side-on adsorption of these systems was not considered in the following study (TM@Sb, TM = Co, Ni, Rh, Ru, Pd, Pt, Ir, Lu). In detail, most systems prefer end on adsorption. It is worth noting that, different from the previous research conclusions that is the $N_2$ is still energetically favorable to be adsorbed in a standing-on pattern, some systems of this work show the comparative adsorption results between end on and side on in a same extent, such as TM@Sb (TM = Ta, Lu, Co, Rh, Ir, Ni), which significantly broadens the application of NRR.[42, 43] According criterion 2, 15 and 9 TM@Sb of end-on and side-on configurations with $\Delta G_{*N2}$ values between -1.06 to -0.30 eV are singled out, respectively. Hence, introducing appropriate doped atoms into Sb monolayer can significantly improve the activity of adsorption site and effectively activate $N_2$. In terms of bond length, the N-N bond of these systems are elongated compared to the free $N_2$ molecules (1.12 Å).

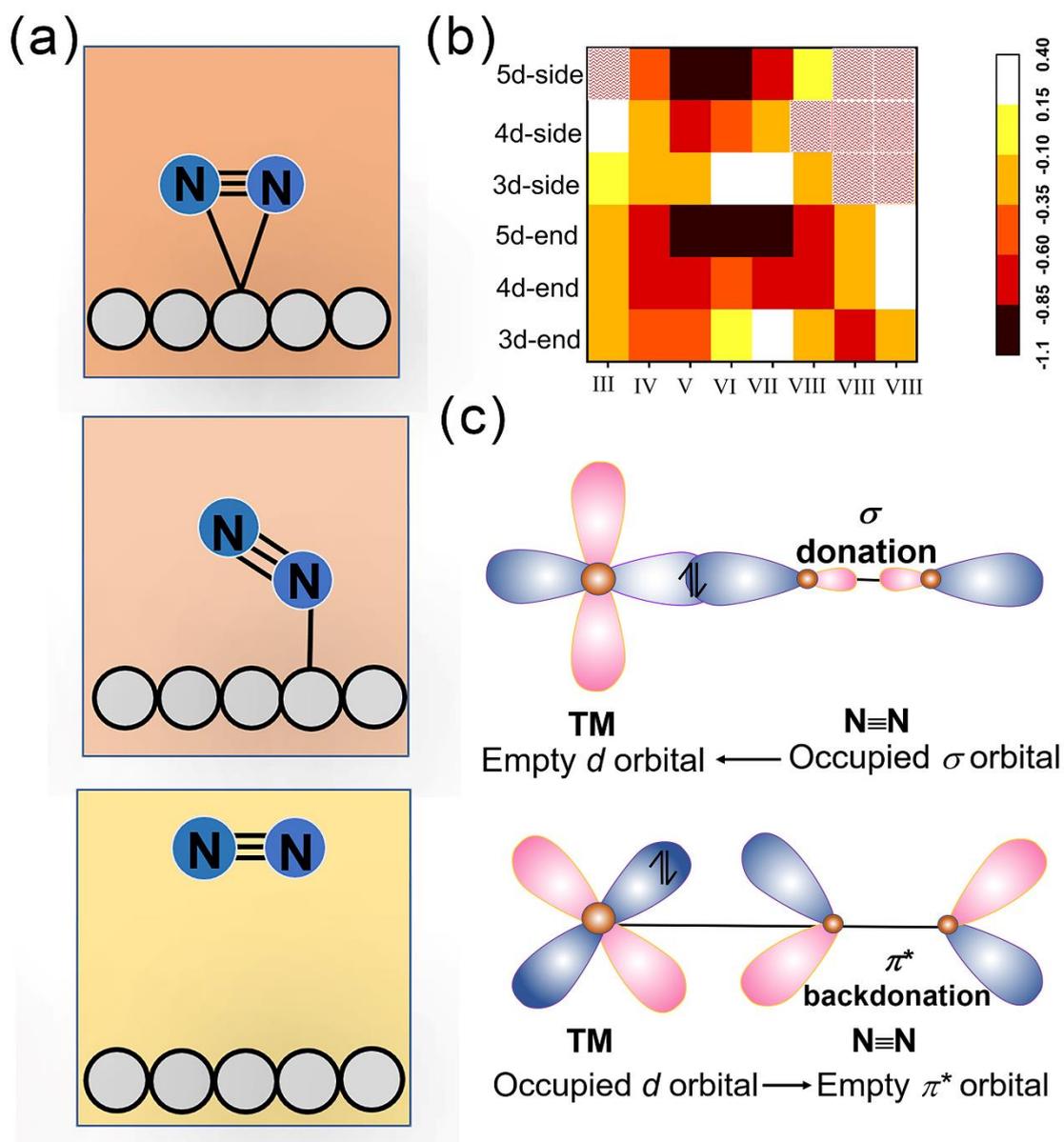

Figure 3. (a) Three possible configurations of N$_2$ on TM@Sb surfaces. (b) Gibbs free energies of *N$_2$ on TM@Sb via both end-on and side-on patterns, where the white indicate that N$_2$ cannot be adsorbed on TM@Sb via this pattern. Black and orange colors present strong and weak binding strengths, respectively. The stripe pattern represents the conversion of side on to end on configuration. (c) The schematic diagrams of N$_2$ binding to single-metal.

As shown in Figure 3c, a single metal site is able to activate the N$_2$ molecule effectively due to the advantage of the interplay between empty and partially filled *d* orbitals of TM atoms, which is consistent with the electron accepting and feeding back mechanism. Notice that the partially occupied $d$ orbital of the metal atom is able to donate electrons to the antibonding orbital of $N_2$, while the lone pair of electrons of $N_2$ is also donated back to the empty orbitals of the metal atom. Here, Figure 4 shows the optimized configurations and charge density difference of Re@Sb surfaces chemisorbed by $N_2$, and other systems are placed in Figure S5—S8. It can be seen that the $N_2$ molecules adsorbed on the catalyst surface gain electrons and thus accumulate a large amount of negative charge. The calculated Bard charge also verifies the result. It is found that 0.26e to 0.75e are transferred from the $d$ orbital of the central TM to the antibonding orbital of the $N_2$ molecule. The effective charge transfer is also the underlying reason for the high activation of the $N_2$ molecule and the significant lengthening of the N-N bond.

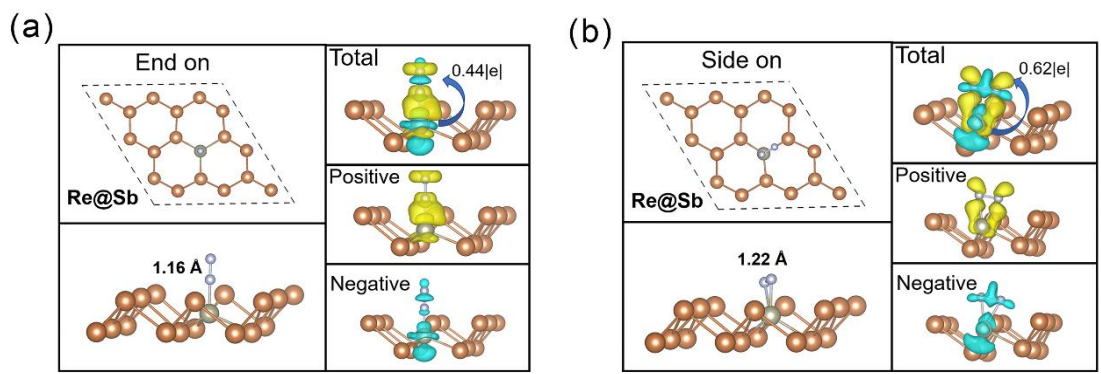

Figure 4. The optimized adsorption configurations and charge density differences of $N_2$ chemisorbed on Re@Sb surfaces. The charge depletion and accumulation were depicted by cyan and yellow, respectively. The isosurface value is 0.002 e/Å$^3$. (a) via end-on configuration (b) via side on configuration.

In order to gain a deeper understanding of the chemisorption process of $N_2$ molecules and its principles, we performed the partial density of states (PDOS) and the crystal orbital overlap populations (COOP) analyses of Re@Sb in Figure 5. Similarly, Re@Sb is used as a representative example here, and other systems are shown in Figure S9—S11. Observing the molecular orbitals of free $N_2$, it can be found that the great advantage of TM@Sb activation of $N_2$ comes from the fact that the occupied $d$ orbitals of TM feedback electrons to the $2\pi^*$ orbitals of $N_2$, which makes the partially occupied $2\pi^*$ orbitals closer to the Fermi level. Therefore, it can be concluded that the strong $d$-$2\pi^*$ coupling can effectively activate the adsorbed $N_2$. This also prepares for the subsequent hydrogenation.

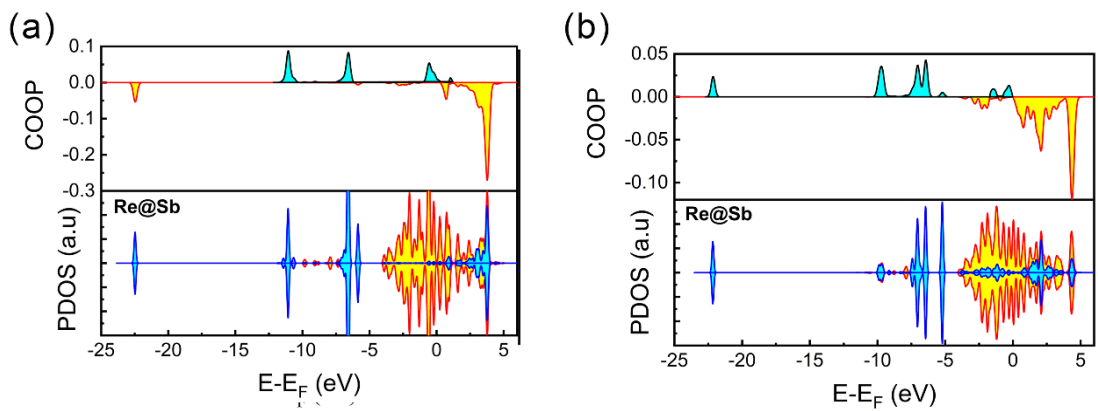

Figure 5. (a, b) The computed partial density of states (PDOS) and the crystal orbital overlap populations (COOPs) of $N_2$ on Re@Sb surfaces via end-on configuration and via side on configuration. The bonding and antibonding states in COOP are depicted by cyan and yellow, respectively.

### 3.3 The screening and NRR performance of SACs.

Next, we further analysis the $\Delta G_{*N2-N2H}$ of TM@Sb based the first hydrogenation step (i.e., $*N_2 + H^+ + e^- \rightarrow *N_2H$) via end-on and side-on patterns. As shown in Figure 6. there are 6 and 9 kinds of SCAs via end-on and side-on patterns are excluded to efficiently catalyze $N_2$ into $N_2H$. Next, the end-on configurations (TM@Sb, TM = Nb, Mo, Ta, W, Re, Os) and side-on configurations (TM@Sb, Ti, Zr, Nb, Mo, Tc, Hf, Ta, W, Re) are further screened out as promising candidates for subsequent calculations.

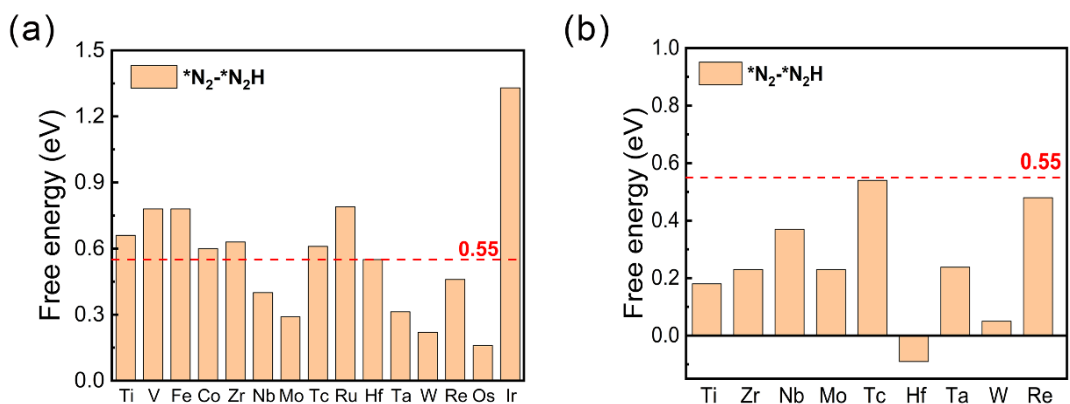

Figure 6. The Gibbs free energy changes of $*N_2 + H^+ + e^- \rightarrow *N_2H$ on TM@Sb. (a)

via end-on patterns. (b) via side-on patterns.

Generally, the NRR has three possible reaction pathways.They are distal, alternating, and enzymatic mechanisms, involving six successive protonation and reduction processes (Scheme 1). For end-on adsorption, the transformation of $N_2$ to $NH_3$ follows the distal or alternate mechanism. Specifically, for the distal path, after $N_2$ adsorption on the catalyst, the distal N atom is first protonated to produce one $NH_3$ molecule and released. Subsequently, the remaining N atom is further protonated until a second $NH_3$ molecule is subsequently produced and desorbed from the surface.[44, 45] For the alternating path, the two N atoms are alternately protonated until the resulting $NH_3$ molecules are desorbed in succession.[44, 45] For the side-on adsorption, the NRR follows an enzymatic reaction in which the two adsorbed N atoms are attacked alternately.

Scheme 1. Schematic of Distal, Alternating and Enzymatic Mechanisms for eNRR.

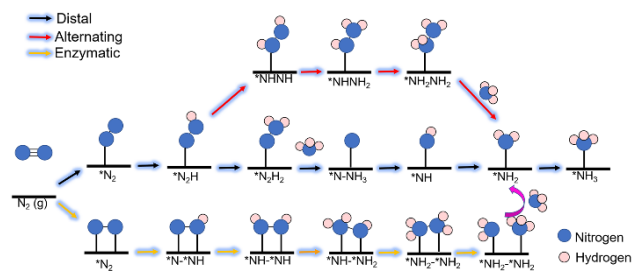

Here, the free energies of the detailed reaction pathways of the 15 TM@Sb were calculated to evaluate their catalytic activities. As a complement, 6 kinds of SCAs are capable of potentially distal and alternating reactions by end-on adsorption, and 9 kinds of SCAs catalyzed enzyme reactions through side-on patterns. For all electrode reactions involving electrons, the effect of the corresponding applied potential (U) is included in the free energy transfer of -neU. A smaller negative applied potential indicates a smaller overpotential. Figure 7(a-c) shows the Gibbs free energy of the overall NRR paths on Re@Sb, and other systems are shown in Figure S16-S21. In addition, the optimized structures of the reaction intermediates on the TM@Sb surface are given in Figures S12-S14 as representative. It is worth mentioning that when Ta@Sb performs NRR via alternating pathway, the intermediate structure of $*NH_2NH_2$ is transformed to *NH-NH$_3$ of distal pathway (Figure S15). Therefore, the subsequent studies do not consider its alternating reaction.

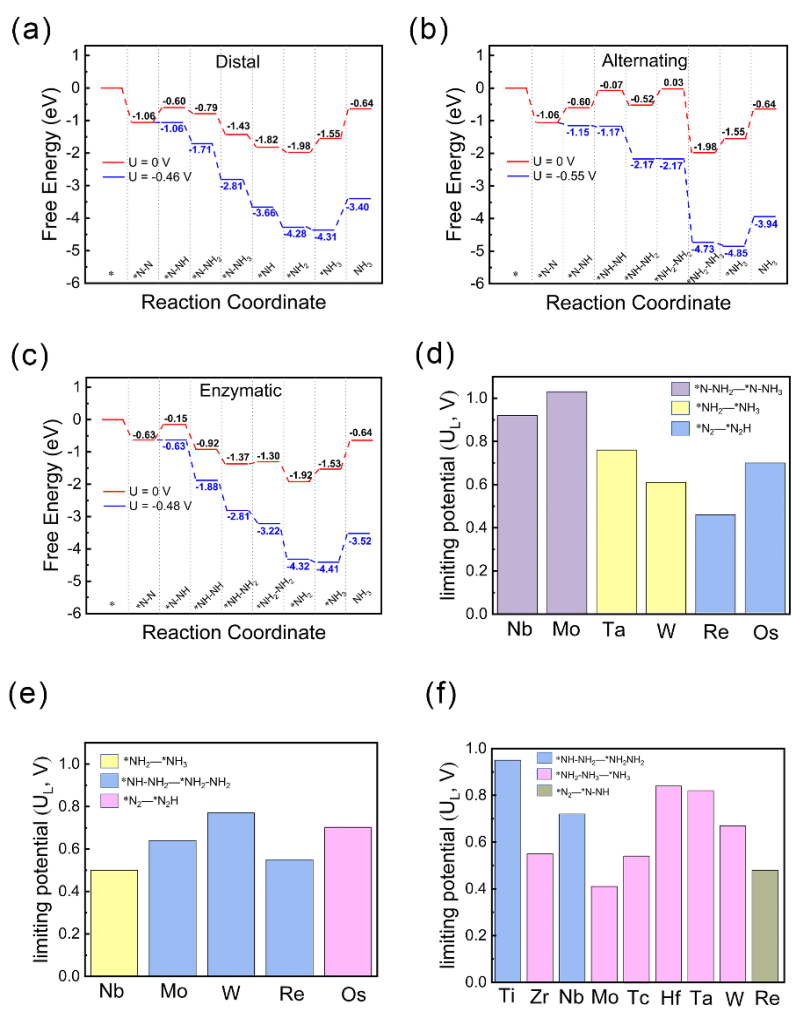

Figure 7. The Free energy diagrams during the NRR on Re@Sb. (a) via distal pathway (b) alternating pathway. (c) enzymatic pathway. The limiting potential (U$_L$) for NRR on the promising TM@Sb candidates. (d, e) via distal and alternating pathway (TM = Nb, Mo, Ta, W, Re and Os). (f) enzymatic pathway (TM = Ti, Zr, Nb, Mo, Tc, Hf, Ta, W and Re).

The potential-determining step (PDS) is the one with the largest change in positive free energy of all the basic steps of the reaction. Particularly, the PDS of possible reaction pathways and the limiting potential (U$_L$) are summarized in Table S1—S3 and Figure 7 (d—f). In order to reveal the reaction mechanism, we further established the activity trend relationship of various SACs. For distal mechanisms, there is an apparent linear relationship between the reduction limiting potential of NRR and the Gibbs free energy of $N_2$ adsorption ($\Delta G_{*N2}$), as shown in Figure 8a. Therefore, the value of $\Delta G_{*N2}$ acts as a descriptor to evaluate the overall NRR activity. To gain insight into the origin of catalytic activity, we further studied the electronic structures of TM@Sb systems. Combined with the analysis in Figure 4 and Figure 5 above, we visually present the orbital interaction diagram in Figure 8d. the $d$ orbitals of the TM atom interact with the $p$ orbitals of N atom of $*N_2$ directly adsorbed on the TM sites and the populations above/below $E_f$ signify the anti-bonding/bonding state. For these doped systems with activated $N_2$, the $d$ orbital of the TM atom interacts with the $p$ orbital of the N atom of $*N_2$ adsorbed directly on the TM sites to form the antibonding and bonding states, located on and below the $E_f$, respectively. This theoretical explanation of orbital interaction has been applied to other catalysis as well.[11, 46]

In common with other catalysts, the different binding abilities of various Sb-based doping systems to $N_2$ can be correlated with the bond order ($p$), defined as follows:

$$p = \frac{n(\sigma) - n(\sigma*)}{2} \quad (6)$$

where n($\sigma$) and n($\sigma$*) are the electron number of $\sigma$ and $\sigma$*, respectively. Figure 8b and 8c shows a linear relation between $p$ and $U_L/\Delta G_{*N2}$. Broadly speaking, the smaller the $p$, the stronger the catalyst binds to $N_2$, the smaller the limiting potential of the whole reaction, and the better the catalytic performance. Establishing such a relationship between catalytic activity and electronic properties could have a profound impact on the development of 2D materials for energy conversion to a large extent.

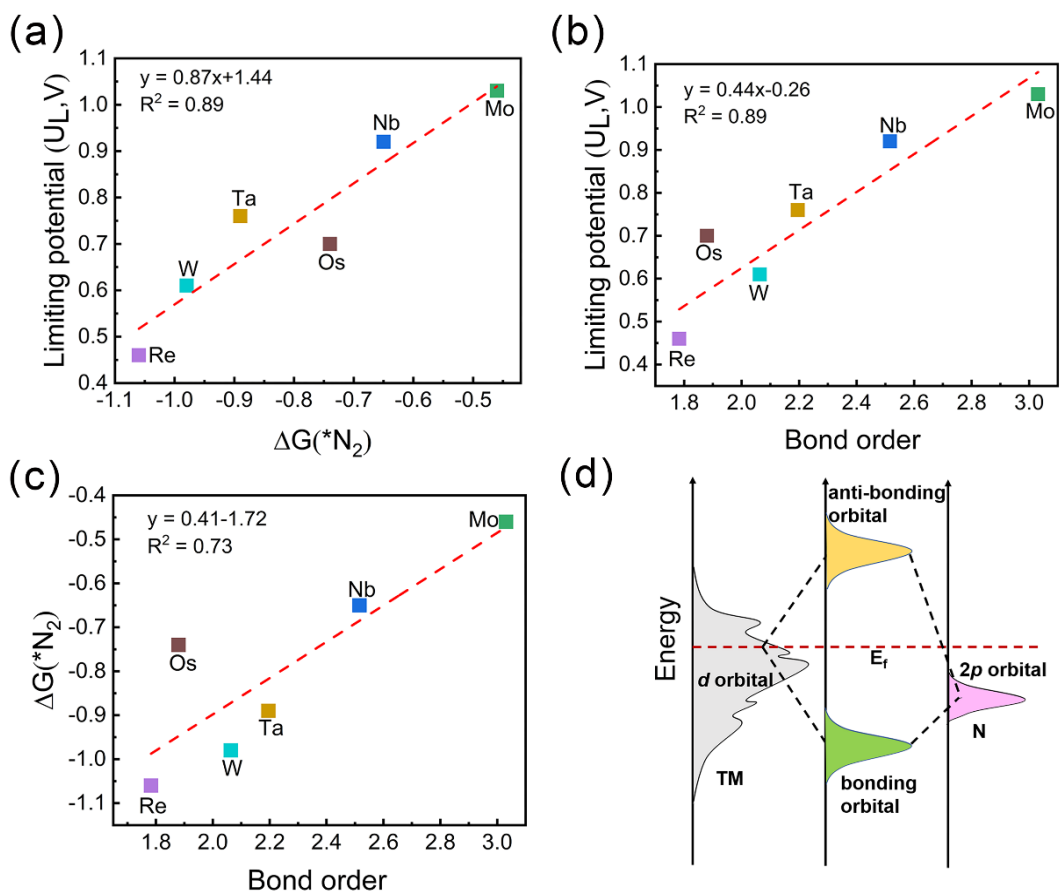

Figure 8. (a) The linear correlations between the adsorption energy of N$_2$* ($\Delta G_{*N2}$) and limiting potential of the NRR by distal mechanisms (U$_L$) on various TM@Sb model surfaces. (b, c) The limiting potential (U$_L$) and $\Delta G_{*N2}$ of various TM@Sb as a function of the bond order (*p*), respectively. (d) Interaction diagram of the 2*p* orbital of N atom of *N$_2$ adsorbates which are directly binding with TM sites and the *d* orbital of the TM center with end-on or end-on configuration.

As shown in Table S2, the limiting step of the alternating reaction of TM@Sb is not always the first and last hydrogenation step, but mostly the fourth hydrogenation step (*NH-NH$_2$ + H$^+$ + e$^-$ → *NH$_2$-NH$_2$). Furthermore, the relationship between the adsorption strength of *NH$_2$-NH$_2$ species ($\Delta G_{NH2NH2*}$) and the NRR activity (U$_L$) of alternating mechanisms was investigated, which had a significant linear relationship (R$^2$ = 0.78) (Figure 9a). According to the relationship between $\Delta G_{NH2NH2*}$ and U$_L$, when the $\Delta G_{NH2NH2*}$ approaches 0.46 eV, the U$_L$ has the lowest value, which is well consistent with the free energy diagram in Figure S16. For the enzymatic mechanisms, it was found that the $U_L$ is related to number of valence electrons and electronegativity of TM active site. As shown in Figure 9(b), the descriptor ($\varphi = \frac{n}{\sqrt{E_{TM}}}$), containing the electronegativity ($E_{TM}$) about elemental physical parameters of the dopant, and the total number of electrons in the valence orbital (*n*) of TM, exhibits a negative linear correlation with the limiting potential ($U_L$) of enzymatic mechanisms.

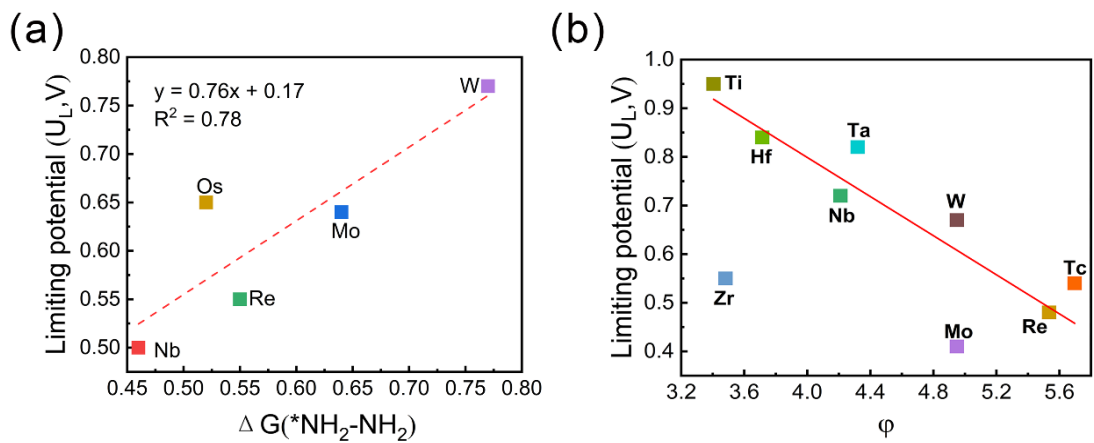

Figure 9. (a) The linear correlations between the Gibbs free energy of *NH$_2$NH$_2$ ($\Delta G$(*NH$_2$NH$_2$)) and limiting potential of the NRR by alternating mechanisms ($U_L$). (b) The scaling relationships between the limiting potential of the NRR by enzymatic mechanisms ($U_L$) and of $\varphi$.

Overall, based on criteria 4, we further screened the PDS of the most promising TM@Sb catalysts, with 0.55 as the benchmark for further narrowing the screening range. Notably, in combination with figure7 (d—f) and criteria 4, there are Re@Sb, Nb@Sb and TM@Sb (TM = Zr, Mo, Tc, Re) available for reactions distal, alternating pathway and enzymatic pathway, respectively, which will be considered for further study.

### 3.4 The competition between the NRR and HER of SACs.

In order to achieve higher Faraday efficiency of NRR ($f_{NRR}$), the catalyst should be able to effectively inhibit HER in addition to stability and activity screening. Therefore, selectivity of the catalyst was finally screened. In this case, the Gibbs free energy difference ($\Delta G = \Delta G_{*N2} - \Delta G_{*H}$) between *H ($\Delta G_{*H}$) and *N$_2$ ($\Delta G_{*N2}$) was calculated to evaluate the actual catalytic activity of different catalysts. When the $\Delta G$ value is negative, it means that the adsorption of N$_2$ on TM@Sb is more favorable than the adsorption of H. As shown in Figure 10 (a) and Figure S22, only Zr@Sb catalysts with H adsorption dominate the candidates screened by the above screening criteria. After considering the limiting potential (U$_L$) between NRR and HER, there are only Re@Sb and Tc@Sb ideal catalysts that meet the proposed screening criteria, as shown in Figure 10 (b). Among them, based on ideal catalytic performance, Re@Sb can achieve NRR under three mechanisms, Tc@Sb is conducive to NRR under enzymatic mechanism. The NRR catalytic performance of these potential ideal catalysts is described in the next section. To summarize, among the 15 TM@Sb, we identified that Re@Sb and Tc@Sb exhibit higher activity toward the NRR. Besides, the Faraday efficiency of the NRR ($f_{NRR}$) was considered. For Re@Sb, the limiting-potential step of HER is 0.67 eV and NRR is 0.46, 0.55, 0.48 eV via distal, alternating and enzymatic pathways, respectively. And the corresponding limiting potentials of HER and NRR for Tc@Sb are 0.24 and 0.14 eV. Moreover, the $f_{NRR}$ values of Re@Sb and Tc@Sb were about 100% at ambient temperature, suggesting that considerable NRR selectivity was exhibited by inhibiting competitive HER.

It is worth noting that unlike NRR process occurs under strong alkaline conditions and thermal catalysis, the desorption of NH$_3$* is essential in the overall process.[47] After overcoming the positive $\Delta G$ of 0.84 eV and 0.87 eV, NH$_3$ will desorb from the surface for Re@Sb via distal, alternating and enzymatic pathways, respectively (Figure 7), and the Gibbs free energy of NH$_3$ desorption on Tc@Sb is 0.76 eV. The NRR is a six-electron reduction process, while the last step is independent of the overpotential since it does not involve hydrogenation. In order to ensure efficient catalyst utilization, the generated NH$_3$ should be made immediately desorbable from the catalyst by applying a potential during the experiment. Considering practical applications, the lower the $\Delta G$ in the last step, the better. On the other hand, the adsorbed *NH$_3$ may be further protonated to form NH$_4^+$, which is not considered in this work since it requires the construction of a more detailed model of dissolved $NH_4.^+$[48]

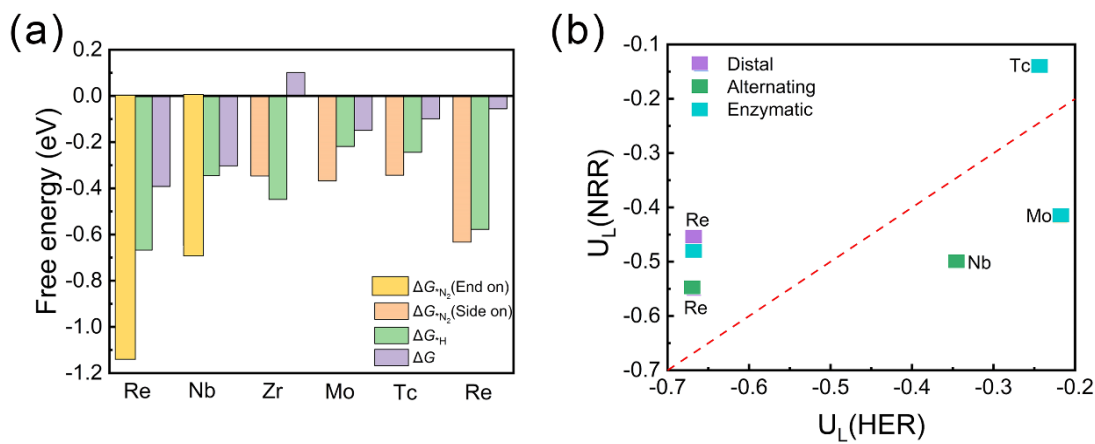

Figure 10. (a) The Gibbs free energy of $N_2$ ($\Delta G_{*N2}$) and H adsorption ($\Delta G_{*H}$) on TM@Sb as well as the difference ($\Delta G = \Delta G^*_{N2} - \Delta G^*_H$) between them via end-on patterns and side-on patterns. (b) The limiting potentials for NRR ($U_L(NRR)$) and HER ($U_L(HER)$) illustrating the NRR selectivity of TM@Sb via Distal pathway, Alternating pathway and Enzymatic pathway.

## 3.5 Origin of reactivity and Potential application of ideal catalysts.

Subsequently, the charge changes of all possible pathways were considered to analyze main source of NRR performance. First, the charge variations of Re@Sb and Tc@Sb were illustrated by the Bader charge, as displayed in Figure 11, where step 0 indicates the charge transfer of the catalyst after $N_2$ adsorption. Each intermediate is divided into three moieties: the defective Sb substrate (moiety 1), TM atom (moiety 2), and the adsorbed $N_xH_y$ groups (moiety 3). Apparently, for all pathways, the charge changes of moiety 2 and moiety 3 show opposite trends. In NRR, the TM sites participate in the reaction or act only as electron transporters. For example, in the fourth step of the Re@Sb enzymatic pathway, the adsorbed *$NH_2$-$NH_2$ acquires electrons ~ 0.2 e from defective Sb substrates via TM. This indicates that the substrate is considered as an electron reservoir and the transition metals can be regarded as electron transmitter that transfer charges between moiety 1 and moiety 3, facilitating the NRR reaction. In parallel, we also investigated the bond lengths along different mechanisms. The variation of bond length for TM–N ($d_{TM-N}$) and N–N ($d_{N-N}$) during reaction, as displayed in Figure S23 and inset. Apparently, $d_{N-N}$ shows a monotonic increasing trend before the release of the first $NH_3$ molecule, indicating that $N_2$ molecules can be efficiently activated on Re@Sb and Tc@Sb.

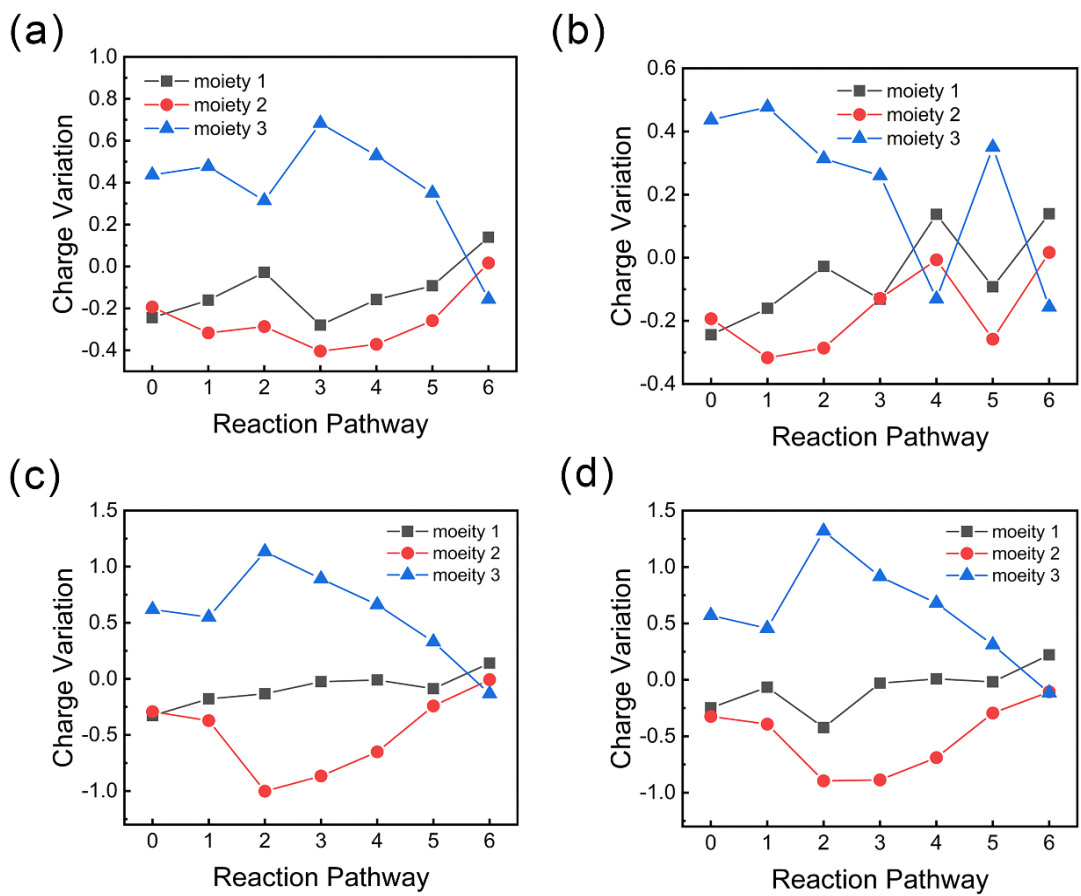

Figure 11. (a—c) The charge variations of the three moieties on Re@Sb along the distal, alternating and enzymatic pathways. (d) The charge variations of the three moieties on Tc@Sb along the enzymatic pathway.

In the end, the electronic properties of the ideal catalysts were further analyzed, which are important for activating $N_2$ molecules and facilitating charge transfer leading to efficient electrocatalytic processes.[49] In addition, the band structures of Re@Sb and Tc@Sb were calculated taking into account their electronic conductivity. The results show that these materials exhibit metallic properties, which are significantly different from the semiconductor properties and the band gap of 0.53 eV presented by the pristine Sb monolayer (Figure S24). Thus, the introduction of Re or Tc elements significantly enhances the electronic conductivity of the Sb monolayer, facilitating electrocatalytic nitrogen reduction. Furthermore, we also calculate the I-V curve of the ideal catalysts at bias voltage up to 2 V, as shown in Figure 12a. To clarity, at a bias voltage V = 1.4 V, the Sb based catalyst begins to produce a current. Compared with pristine Sb, the current of Re@Sb and Tc@Sb increases obviously with the increase of voltage and reaches a fairly reasonable size of current between V= 1.4 V and V= 2.0 V. This further verifies that the selected ideal catalyst has good electronic conductivity. Here, curve(I) of Re@Sb and Tc@Sb is taken as the sum of the spin up current and spin down current, and their spin-dependent I-V curves are shown in detail in Figure S25. In order to obtain a quantitative result, for example, we then explain their current change by analyzing the corresponding transmission spectrum with bias of 1.8 V, as shown in Figure 12(b) and (c). For pristine Sb, the transmission coefficients of the spin-up and -down channels calculated are exactly the same, so there is only one curve (black dotted line) in Figure 12(b) and (c). For the doped system, the two spin channels have two transmission curves. And the phenomenon of nonequilibrium quantum transport can be discussed by analyzing the transmission spectra under finite bias V, and the current can be obtained by integrating against the bias window $-V/2 \leq E \leq +V/2$. Indeed, at the bias window $-0.9\ V \leq E \leq +0.9\ V$, the transmission coefficient of the Re@Sb and Tc@Sb systems is relatively larger than that of the pristine Sb, indicating that the current is relatively strong, which explains the behavior in Figure 12a.

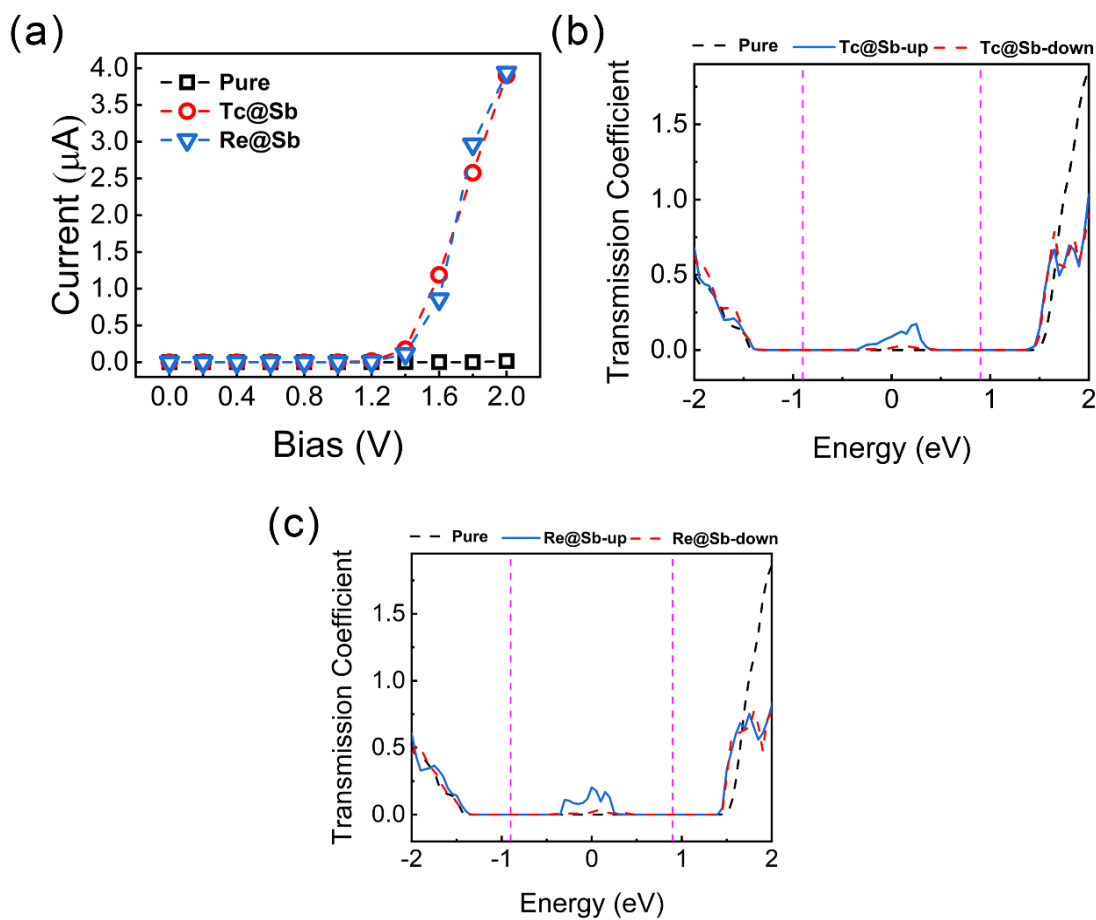

Figure 12. (a) The calculated I-V characteristics of the current I versus the bias voltage V. (b, c) The transmission spectrum of the models under 1.8 V biases.

The excellent electronic conductivity gives Re@Sb and Tc@Sb excellent catalytic performance in NRR. Given that the NRR process usually occurs in aqueous solutions and the Re or Tc sites in TM@Sb are likely to be occupied by water molecules in the environment, the adsorption of $H_2O$ on the Re site to was also evaluated to predict the stability in practical applications. In addition, the optimized adsorption configuration for $H_2O$ in Figure S26. By calculation, the Gibbs free energy of $H_2O$ at the Re site is -0.29 eV, compared to the value of $N_2$ at Re@Sb through the end or side conformation (-1.06 eV and -0.54 eV), which indicates that $H_2O$ does not pose competition during NRR. The Gibbs free energy of $H_2O$ on Tc site is -0.25 eV and that at Side on configuration is -0.34 eV. These results indicate that $N_2$ can selectively adsorb to the central Re or Tc site, offering a great advantage over its counterparts. In conclusion, Re@Sb and Tc@Sb are indeed a potentially ideal catalyst for screening.

## 4. Conclusion

In summary, we systematically explored the effects of TM embedded on the NRR catalytic activity of $\beta$-Sb monolayer by performing high-throughput screening of DFT computations. This screening principle allows efficient design of catalysts and saves significant computational resources. Our results showed that the introduction of suitable dopant can effectively activated $N_2$ molecule, thus improving its NRR activity. Besides, combination of the selectivity, stability and durability analyses as well as the synthetic accessibility supports the great potential of Re@Sb and Tc@Sb as NRR electrocatalyst. For the whole reaction pathway, Re@Sb enables distal, alternating and enzymatic pathways with limit potentials of -0.46 -0.55 and -0.48 V, respectively, while Tc@Sb enables the enzymatic pathway with a limit potential of -0.14 V. In particular, the NRR activity descriptor for the TM@Sb system was proposed. For the exploration of two-dimensional SACs in NRR and other electrochemical reactions, our work will stimulate more experimental and theoretical further efforts.

## Conflicts of interest

There are no conflicts to declare.

## Acknowledgements

This work was supported by the National Natural Science Foundation of China (51972227) and (11804023), and the Natural Science Foundation of Tianjin (18JCQNJC02700).